\documentclass[11pt]{article}

\usepackage[pdftex]{color}
\usepackage{graphicx}
\usepackage{indentfirst}
\usepackage{latexsym}
\usepackage{amsfonts}
\def\noi{\noindent}

\begin{document}
% ----------------------------------------------------------------

\title{%\huge {
%\textbf{
A framework to streamline the process of systems modeling
%} %}
\author{$^1$Elias Carvalho, MSc
%\thanks{\noi Universidade Estadual de Maringa, Maringa, Brazil.}
\and $^2$Luciano Andrade, RN, MSc
%\thanks{\noi Universidade Estadual do Oeste do Parana, Foz do Iguassu, Brazil.}
\and $^3$Ricardo Chaim, PhD  
%\thanks {\noi  Universidade de Brasilia, Brasilia, Brazil} 
\and $^4$Ricardo Pietrobon, MD, PhD, MBA
%\thanks{\noi  Duke University Health System, Durham, USA.}%\thanks {Partially supported by CONICET (PIP 4463/96), Universidad de La Plata (UNLP 11 X472) and
%ANPCYT (PICT03-09521).}
}
\date{}
}
\maketitle

\noi $^1$Universidade Estadual de Maringa,  Maringa, Brazil.

\noi e-mail: ecacarva@uem.br 

\medskip

\noi $^2$Universidade Estadual do Oeste do Parana, Foz do Iguassu, Brazil.

\noi e-mail: luciano.andrade@unioeste.br

\medskip

\noi $^3$Universidade de Brasilia, Brasilia, Brazil.

\noi e-mail: ricardoc@unb.br

\medskip

\noi $^4$Duke University Health System, Durham, USA.

\noi e-mail: rpietro@duke.edu

\begin{abstract}
Although simulation represents a major advance in the understanding of problems in complex systems, the field currently does not has standards in place that would guide the reporting of the data underlying each model, the process for model validation.  This lack of common platforms significantly decreases the attractiveness of the field in that comparison across models is difficult, and peer evaluation of whether models should be trusted and used in practice is severely limited.  This article reports on a series of concepts that could serve as the basis for an initial discussion regarding potential platforms wiht the simulation community.  The document covers a proposed platform for research question formulation, literature review, data collection, model analysis, and manuscript writing.  Considerations are then made to counter the prediction that raising quality levels would lead to a decreasing rate in expansion for the field.

\


\end{abstract}

\newpage

\section{Introduction}

Simulation represents a major analytical advance in a number of scientific fields, being used for both prediction as well as understanding of problems in systems demonstrating complex behavior \cite{GR2009}. Used in multiple fields, from economics to medicine, simulation frequently generates substantial insights regarding systems that otherwise would be only partially understood through piecemeal analytical approaches that are incapable of apprehending the multiple interactions, feedback loops, and unintended consequences associated with complex environments.  Despite its significant methodological achievements, simulation models are often plagued by a development methodology that is unclear, based on data sources that are undefined, and drawing relations to the real world that are questionable.
	
To put the simulation of complex systems in perspective, a contrast should be established with previous methods and their limitations.  These previous methods include classical techniques of experimentation that can only control for a few variables at a time, deterministic solutions that require an exact analytical solution, and classical statistical models that are primarily bound to quantitative data sets. In contrast, simulation allows for the inclusion of quantitative and qualitative data, complex interactions among multiple variables, and model manipulation in a variety of ways that allow for prediction, better understanding of the model in relation to the problem being studied, and the analysis of multiple hypothetical interventions.  All of these activities are, however, pending not only on how the model was originally constructed (i.e., which sources of data and information were considered in its formulation), but also how the model has been validated (i.e., compared to problems in the environment where the problem originally occurred).  Currently, a variety of models in the literature lack any use of platforms to streamlined how these models are built and validated \cite{Rand2006}.

The objective of this proposal is therefore to review the literature and propose an initial set of platforms for the creation of simulation models, common to System Dynamics (SD).  These platforms are divided into research question formulation, literature review, data collection, modeling process, and manuscript reporting, each one covered in the following sections.

\section{Research Question Formulation}

Based on a compilation of previous references from SD literature \cite{PenaLyneis2009}, we propose that the process of research question formulation be composed of three main section.  Namely, these sections are (1) Hypothesized conclusion, (2) Variable selection, and (3) Mock results.  

\subsection{Problem definition through a "hypothesized conclusion"}

Hypothesized conclusions follow the Feinsteinian tradition \cite{Feinstein1987} of stating the supposed conclusion of a study in order to extract the real intent of the researcher stating the problem to be resolved.

According to Feinstein, whenever a researcher attempts to formulate the problem to be solved in terms of hypothesis or some other vague statement, their ultimate purpose and belief of what they would like the model to ultimately solve or display is hidden.  Another important aspect in this process is that the model has to be based on a problem rather than on an attempt to mimic a system.  Because systems are complex and can be perceived from multiple perspectives \cite{Giere2006}, a system could be modeled in a number of different ways depending on the problem that the modeler is attempting to approach.  

\subsection{Variable selection}

Variable selection is important in that it will assist in the creation of an outline of the SD structure. Variable selection is primarily determined by a combination of information from the literature as well as quantiative and qualitative data, which are treated in more detail in subsequent sections (sections 3 and 4, respectively).  Finally, variable selection is closely related to the hypothesized conclusion, in that it should contain all variables required to address the problem, but probably no additional variables in order to decrease model complexity and unnecessary work.

The variable selection process involves the compilation of factors that are believed to be implicated in the problem to be modeled.  For SD models, this list can later be organized in a causal loop diagram demonstrating the feedback loops involved in their interaction as demonstrated in the "Standard Method" outlined by Lyneis \cite {PenaLyneis2009}. 

\subsection{Mock results}

Mock results are important to align the problem statement outlined in the hypothesized conclusion with the list of variables.  Briefly, when the mock results do not clearly present an answer to the stated problem or when it is based on variables that have not been previously listed, the research question formulation process is inconsistent and should be revised.  Mock results take a number of forms outlined in the following sections.

\subsubsection{Reference modes}
\begin{figure} [htbp]
   \centering
   \includegraphics[width=0.70\textwidth]{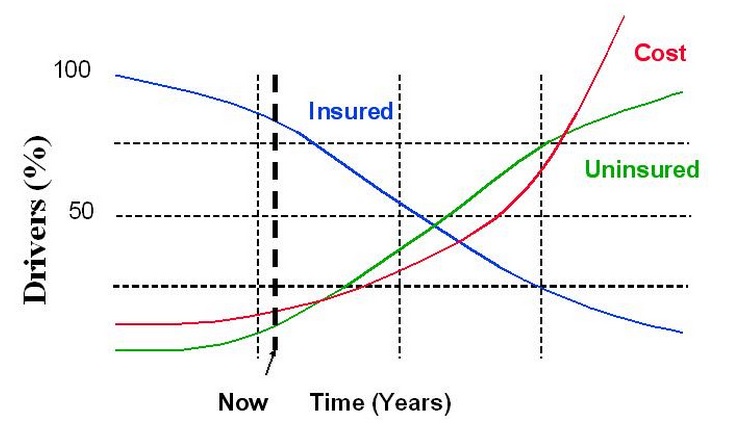}
  \caption{Example of reference}
  \label{fig:figart}
\end{figure}

Reference modes constitute a graphical representation of what modelers believe the system response will be over time \cite {PenaLyneis2009}.  This representation usually contains best and worst case scenarios. For example, in the case of a citation network, a high-impact article is expected to obey a power-law up to the point when that citation ages, when its probability of being cited starts declining (Figure \ref{fig:figart}).

\subsubsection {Selection of archetypes and template models}

Once the reference modes are stated, the next step is to search model libraries for pre-existing models or archetypes that can provide structural templates that mimic the expected system behavior.  This process, sometimes called dynamic hypothesis, represents the association between model structure and expected system behavior.  This phase can be compared to the sequence involved in statistical modeling where a family of distributions might be initially chosen to model a given process, the fit of this distribution family to the actual data being later validated.  Similar to the statistical modeling process, initial assumptions can later be replaced, and this is often the case with archetypes.  In sum, they function as initial templates rather than determine how the system should be modeled.

SD archetypes have been portrayed as the structural basis determining common types of system behavior encountered in complex models \cite{Wolstenholme2003};\cite{Wolstenholme2004}; \cite{Braun2002};\cite {systemdynamicsArchetypes};\cite {systemsthinking}.  When some of the expected behavior determined by reference modes is mapped to a certain archetype, that archetype usually provides insight into the model structure.  This statement should not mean that the structure of the model should be driven by the archetype, but simply that archetypes can assist in   providing ideas regarding potential mechanisms that influence the observed behavior.  An additional resource is the concept of molecules \cite{vensimmolecule},  initially developed by Jim Hines \cite{Hines1983}.  Molecules constitute the building blocks of SD models or small templates that can be assembled and customized to create larger models.

The models can then be used, in modules or as a whole, as the basis for modeling the new problem. Libraries for SD models are somewhat limited, although the public library from Tom Fiddaman \cite{metasd} is available, also including personal libraries such as Jim Thompson's personal library for healthcare models (Jim Thompson, personal communication).  

\subsubsection {Knowledge engineering and model libraries}
Although existing libraries represent an important resource, the navigation through their content presents significant limitations.  Libraries usually label and describe their models focusing on their main content area (e.g., an emergency room, a fishery model), whereas modelers searching for model templates are usually interested in models with certain behaviors or structural characteristics.  A potential solution for this problem would be the development of a simple computational ontology \cite{ALLEMANGHENDLER2008}  allowing modelers to perform rich semantic search strategies to locate the best possible model to fit their purposes.  This ontology would contain elements describing four main model characteristics, namely, content area, structure, behavior, and model quality.  While in this article we will not focus on content area, likely organized around a folksonomy \cite{WAL2008}, structure and behavior (requiring a group of experts proposing an initial classification validated through the application to existing libraries), the issue of model quality is central and will therefore be more explored in this article.

A wide variation in model quality variation is pervasive in much of the computer simulation literature.  Two essential areas are frequently unclear whenever a model is built, analyzed and reported:  (1) Data sources used to build the model and (2) How the model is validated or, stated in a different manner, how its behavior relates to the environment where the problem is located.  These two areas are addressed in the following sections.

\

\noindent \textbf{Data sources for baseline simulation models}

\

The proposed classification for data sources in baseline models assumes that the data should be classified in three categories:  Quantitative, qualitative, and personal experience (Table 1). 

\begin{figure}[htbp]
	\centering
		\includegraphics[width=0.95\textwidth]{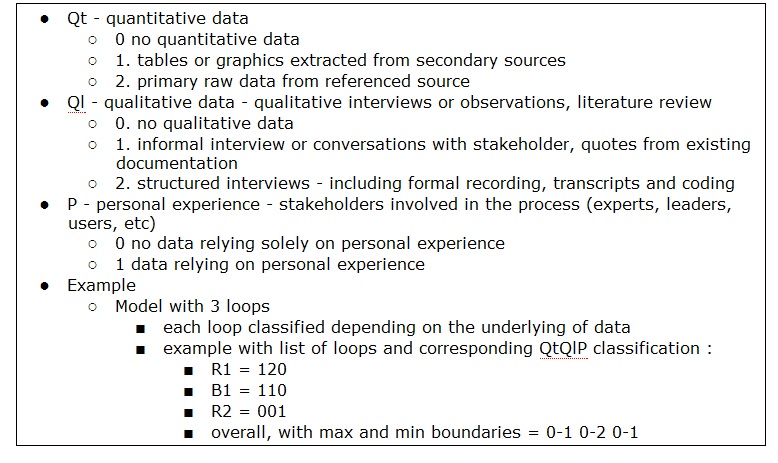}
	\caption{Data source classification for baseline models}
	\label{fig:quadro1}
\end{figure}

For SD models each loop is classified separatel. This proposed model should undergo extensive validation using existing libraries before it can be presented as an initial working version.  Of importance, this classification does not stand for a quality judgment or criteria for selecting one model over another.  It also has no implied assumptions on whether qualitative data might be superior to quantitative or vice-versa.  It simply clarifies where exactly the data and information underlying the model comes from, thus allowing readers to judge for themselves on whether they should or not trust model results. 

The second point where further quality assurance for the model library is required is regarding model validation or how the baseline model corresponds to the behavior presented in the environment where it is located.  A simplified classification for validation is proposed in Table 2. 

\begin{figure}[htbp]
	\centering
		\includegraphics[width=0.95\textwidth]{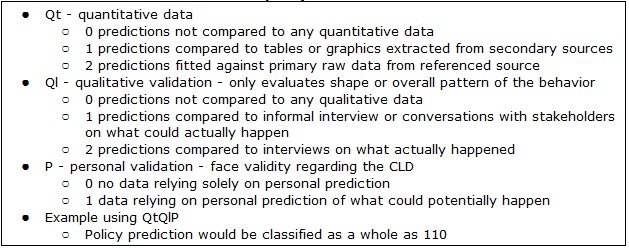}
	\caption{Validation for baseline and policy models }
	\label{fig:quadro2}
\end{figure}

Additional comments on model validation will be made in section 5.2.

\

\noindent \textbf{Model library governance}

\
Governance for this library will include minimal quality criteria for model inclusion as well as incentives for participation.  The latter could include, for example, the provision of references for contributed models that can then be used for citation in articles, curricula, funding proposals, among others.  This incentive framework has been successfully implemented in other projects involving data and model contribution such as Dataverse \cite{thedata}.

\subsubsection {Momentum policies}
Momentum policy \cite{PenaLyneis2009} is a term used within the SD community to represent proposed solutions to the problem being modeled, based on insights obtained from the baseline model.  The potential solutions are implemented in the baseline model, observing whether its behavior changes in the direction expected toward the improvement of the proposed problem.  Similar to the issue of data quality for the baseline model, the description of momentum policies in the simulation literature is frequently unclear, and so we propose that the validation of momentum policies be performed according to the same classification proposed in Table 2, pending further refinement by peers in the SD communities.

\section {Literature review}

\subsection {Content area}
Protocols on streamlined methods to conduct literature review have been prepared by the Research on Research Group at Duke University on how to conduct a literature review \cite{researchonresearch}.  These platforms are related to the research question formulation as well as data collection for modeling \cite{researchonresearch113}.  Briefly, the method of Literature Matrices \cite{PIETROBON2004} is proposed, with articles being grouped in clusters associated with loops for SD.  

\subsection {Simulation modeling}
A literature review for  SD modeling is facilitated by the search of standard sources of information.  These include the bibliographic compiled and constantly updated by the System Dynamics Society \cite{systemdynamics-biblio}, as well as list of journals and books on systems modeling currently in preparation by the Research on Research (RoR) group at Duke University.  In combination with the model library described in section 2.3.3 and mentorship during distance research coaching programs, these resources are expected to improve the overall quality and productivity of the models according to a platform agreed upon by the simulation community.

\section {Data collection}

\subsection {Qualitative data} 
Methods for literature review as a source of qualitative information for simulation models were addressed in section 3.  Also, previous literature has been written on optimal methods to conduct qualitative studies to interview stakeholders (experts, users, among others) providing information that can contribute to the simulation model.  This literature has focused on both qualitative methods in general \cite{Yin2002};\cite{Marshall2006}; \cite{Silverman2004}  as well as specific literature for simulation modeling \cite{Laws2004}; \cite{Luna-Reyes2003}.  Rather than claiming for the superiority of one method vs. another, a platform will simply offer a bibliography on the topic so that modelers can cite their method of choice.  More important than pre-judging which method is superior in each situation, appropriate documentation describing which method was used is essential so that readers can judge whether they should believe or not in the inferences made from the model based on the quality of the data. In situations where the literature is either non-existing or of poor quality, a Delphi method \cite{LINSTONEandTUROFF2002} can be used to obtain expert opinion as the basis for the model. Almost invariably Delphi methods will be the most commonly used methods to define the overall structure of the model as this information is not available from the literature unless previous SD models might already be available from the literature. In conducting a Delphi method, it is important that its methods be reproducible. A dynamic protocol has been developed by the RoR Group that addresses reproducibility within Delphi \cite{SOPDELPHI}.

\subsection {Quantitative data}
The documentation for quantitative methods that might serve as the basis for a model is less well-defined, but we believe that transparency and potential reproducibility continue to be key parameters to be considered.  Therefore, previous models advocating for making data openly available along with the corresponding scripts that gave rise to the analysis used in the results fed into the model should be preferred.  As examples, we cite the sweave \cite{LEISCH2002} and zelig \cite{IMAI2006} libraries for the R statistical language \cite{r-project} as optimized protocols to accomplish this level of transparency. 

\section {Modeling process}
When it comes to the modeling process itself, a platform should be in place that would substantially streamline this process, ultimately improving their quality and productivity.  We only focus on recommendations regarding modularity, distributed versioning, and connection across modeling software packages.

\subsection {Analytical solution / SD}
\subsubsection {Modularity and distributed versioning in model development}
One of the central characteristics of successful distributed software projects is the ability to create modules that can be executed by individual programmers in a streamlined fashion \cite{Moon2000}.  Because systems modeling usually requires the expertise of multiple specialists from a variety of disciplines, and the criticisms of other modelers on how the model should be structured, validated, and reported, model modularity is a key for the expansion of the field.  Modularity in systems modeling can be accomplished by proper documentation of additional structural, behavioral, or validation tasks that should be accomplished, accompanied by the creation of a reliable and easy to use distributed versioning system.

At this point, our group is evaluating the use of Git \cite{GitHub} since it seem to be more amenable to the distributed nature of our network \cite{git} contributing to simulation modeling in comparison with previous tools such as subversion \cite{subversion}.  Ideally, these repository would be made publicly available, so that modelers can download files from regular web pages while those actively working on modifying the model can upload modifications directly to Git.

\subsubsection {Connection across modeling software packages}
One of the major limitations in the field is the lack of widespread interoperability across different modeling software packages.  Ideally, one would like to see the implementation of platforms that allow for at least the connection among three major classes of software packages, including statistical packages (such as the open source package R) \cite{r-project}; SD-specific packages (such as Vensim \cite{vensim}) , Stella/Ithink \cite{iseesystems} , and Powersim \cite{powersim}.

\subsection {Validation}
An extensive literature has been created regarding the multiple challenges associated with the validation (or impossibility of validation) of simulation models using SD.  The literature has plenty of personal opinions, many of them related to the local experience of the modeler with a restricted set of modeling problems while disregarding problems that are in fields that are foreign to them.  Rather than arguing for one approach rather than another, the simple platform displayed in Table 2 (pending modifications and further testing) is proposed, pending further input and refinement by the SD community.  

\section {Manuscript writing}

While platforms in the creation and validation of simulation models are essential, they will realize their full value if they are properly documented in the final article or report.  This section, which proposes a platform for a traditional research paper, relies on information made available by the RoR Group \cite{researchonresearch} from the Duke University Medical Center.  It includes a classical IMReD framework, including Introduction, Methods, Results and Discussion.  For each section, a general discussion of its contents is provided.

\subsection {Introduction}

\subsubsection {Significance}
The first paragraph should state why the topic to be treated in the paper is important. This statement has to convince readers that they should keep on reading the manuscript given its relevance.

\subsubsection {Information gap}
This section should state that, despite being relevant as made clear in the previous section, there is an important information gap in the field. It will be clear in the last section of the introduction that this gap is not any gap, but the gap that will be evaluated by this article. In other words, this section should sell the idea of the importance of your paper to the reader.

\subsubsection {Literature review in support of information gap}
This section reviews the literature on the gap border while pointing to information gap in our current knowledge that will be solved by the article paper. The role of this section is not to review all the knowledge in the field -- that is left for review articles. Instead, the role of this section is to provide readers with the background information necessary to understand the article while constantly emphasizing that no previous article has evaluated the issue that this article will approach. Because the gap is a place where no previous information exists, novice researchers tend to make two mistakes. First, review the literature surrounding the gap while focusing on that literature.  The second common mistake is to attempt to run a literature search focusing only on the gap. Although a search attempting to find information on the gap should certainly be the first step, researchers should not find anything on the gap, otherwise it would not be a gap. A gap, by definition, is an area where no previous information is available. In summary, the literature search to support the existence of the gap is a search focused on the surroundings of the gap, while focusing on making that gap evident to the reader. In other words, researchers should search for the gap and expect to find nothing, and then start expanding their search to the areas surrounding the gap until they find some literature. This ``border'' is then reviewed and criticized for not having yet approached the gap.

\subsubsection {Objective and hypotheses}
The last section concludes the introduction by stating the main objective of the paper and respective hypotheses. Of importance, the objective fills the information gap previously stated in the second section of the introduction.

\subsection {Methods}

\subsubsection {Problem definition}
This section describes the problem to be solved by the model, including how the baseline model can be used to predict the system behavior through policy models.  This section of the article follows the recommendations of section 2.1. 

\subsubsection {Rationale for choice of modeling method}
The model should mimic the conditions of the environment where the problem is situated, explaining why its advantages over other possible research methodologies such as classical experimental methods, deterministic solutions, and classical statistical models. The choice of model (SD) should be described along with a brief explanation of what that modeling option entails.

\subsubsection {Variable selection}
This section describes each of the variables used in the baseline as well as policy models.  For more details, see section 2.2.

\subsubsection {Reference modes}
This section describes the reference modes.  For more details see section 2.3.1.

\subsubsection {Templates and previous examples from model library}
This section describes previous templates or archetypes that might have been used as the basis to address the problem in the current model.  For details, see section 2.3.2.

\subsubsection {Data sources for baseline and policy models}
In this section an explanation of the literature sources (see section 3), sources and methods of qualitative data (see section 4.1), and sources and methods of quantitative modeling process (see section 4.2) are described.  This section includes the classification proposed in Table 1.

\subsubsection {Measures of model validation}
This section determines how the baseline model was comparable to the qualitative and quantitative sources of data as well as which criteria were used to consider the baseline model adequately reproducing the problem in the system being modeled.  This includes a description of momentum policies (see section 2.3.4).  This section should include the classification proposed in Table 2, pending further refinement by peers in the SD community.

\subsubsection {Software}
This section will list software packages and programming languages used in the modeling process.

\subsection {Results}

\subsubsection {Baseline model}
This section presents the overall description of the baseline model. For SD this would include a description of each of the loops represented by causal loop as well as stock and flow diagrams. Source code should be provided in the appendix so that the model can be reproduced. This description is followed by graphics and tables describing model performance metrics at baseline. A description of how the model compares to the problem in the original environment should be provided, preferably with graphics and tables that allow the reader to compare and judge whether model and problem in the original environment are indeed comparable in their behavior and/or predictions.

\subsubsection {Policy models}
This section presents modifications to the baseline model in terms of all variables and structure to the elements of SD. Graphics and tables should be presented comparing multiple policy models, emphasizing in which situations each of the alternative models might be better than others. Information on how these policy models might compare to what would happen for problems in the environment with some of these modifications should be provided. Examples would include partial implementations in similar environments, slightly different implementation in similar environments, or any other data or information that might enhance the readers' ability to judge whether the policy results in the model are indeed likely to result in similar modifications in the environment where the problem is located.

\subsection {Discussion}

\subsubsection {Strengths and summary of main findings}
In this section the main strengths of the model are stated, while summarizing the main results. If possible, it should be stated how the model is unique compared to the remaining literature. The number of main findings is usually restricted to a maximum of four, with a short statement describing each one.

\subsubsection {Discussion of each main model results} 
This section compares each main result of the model (usually three to four) with the literature. First, it should discuss the findings that agree with previous studies, later approaching the model results that disagree with the literature. Attempt to provide explanations on why the disagreement occurred. Disagreements are usually related to a different research methodology and/or a different patient characteristics.  Both qualitative and quantitative aspects of the results should be discussed. This section differs from previous portions of the report in that it allows for more freedom of opinion about interpretation. That said, common sense is always advisable.

\subsubsection {Study limitations}
This section points to the main limitations of your study and what you have done to minimize those limitations (if nothing can be done to minimize them, explain how the limitation didn't have much of an influence on the results). It should also mention analyses that were not performed in the present study and that should be approached in subsequent studies in order to complement the results.

\subsubsection {Conclusions}
This section states the main conclusions based on model findings, including policy implications, potential changes in practice, and future research opportunities.

\section {Some concerns:  Would a platform limit further expansion of the field or even impose limits to its creativity?}
One of the common concerns regarding the implementation of platforms in a scientific field is that they will limit the expansion of the field since the bar will be set too high for newcomers.  Since the inflow of newcomers would be lower, the number of members would decrease over time.  We would argue that this statement oversees that the what attracts newcomers is not that the field has scientific rules that welcome a wide variety of quality levels in modeling studies.  Instead, what would attract newcomers is if suggested platforms were in place so that results can be compared across different models, ultimately leading the field might to accumulate information and later be taken into new, more sophisticated directions.  Although not having any empirical evidence to demonstrate this statement, the concern raised by those concerned with setting a platform seems follows along the SD archetype called "drifting goals" \cite{systems-thinking}, in that not adhering to a platform undermines the intended balance to increase the size and quality of the overall simulation community.

\section {Conclusion}
This article outlines some initial concepts that can be used as a springboard toward the creation of a standardized platform for the design, data instantiation, and reporting of System Dynamics models. We do not cover standards for model dissemination, although we believe that methods such as Mediated Modeling \cite{MedModel}are equally important to ensure that SD models can be used to modify practice rather than simply serve as a theoretical exercise. We hope that some of these ideas can elicit discussion among SD modelers.

%INICIO DA BIBLIOGRAFIA

%FIM DA BIBLIOGRAFIA


\begin{thebibliography}{}

\bibitem{GR2009} Giere R.N.(2009), Is computer simulation changing the face of experimentation? Philosophical Studies, 143, 59-62.

\bibitem{Rand2006} Rand W. and Wilensky U.(2006), Verification and Validation through Replication: A Case Study Using Axelrod and Hammond�s Ethnocentrism Model. Annual Conference of the North American Association for Computational Social and Organizational Sciences.

\bibitem{Feinstein1987} Feinstein A.R. (1987), Clinimetrics, New Haven, Yale University Press.  

\bibitem{PenaLyneis2009} Pena-Mora F. and Lyneis J.M. (2009), System and Project Management, http://www.goodreads.com/book/show/5023938-system-and-project-management.

\bibitem{Giere2006} Giere R.N.(2006), Scientific Perspectivism, Chicago University Press. 

\bibitem{Wolstenholme2003} Wolstenholme E.F.(2003), Towards the definition and use of a core set of archetypal structures in system dynamics. System Dynamics Review, 19, 7-26. 

\bibitem{Wolstenholme2004} Wolstenholme E.F. (2004), Using generic system archetypes to support thinking and modelling. System Dynamics Review, 20, 341-356.

\bibitem{Braun2002}Braun W.(2002), The System Archetypes,  http://wwwu.uni-klu.ac.at/gossimit/pap/sd/wb-sysarch.pdf

\bibitem{systemdynamicsArchetypes} Bellinger G. (2008), Archetypes Systems, http://www.systemdynamics.org/wiki/index.php/System
Archetypes

\bibitem{systemsthinking} Bellinger G. (2004), Archetypes Interaction Structures of the Universe,  http://www.systems-thinking.org/arch/arch.htm

\bibitem{vensimmolecule}Ventana systems inc, http://www.vensim.com/molecule.html 

\bibitem{Hines1983}Hines J.H.(1983), "New Coflow Equations" MIT System Dynamics Group working paper D-3488.

\bibitem{metasd}Tom Fiddaman's System Dynamics Model Library, http://www.metasd.com/models

\bibitem{ALLEMANGHENDLER2008}Allemang D. and Hendler J.A.(2008), Semantic web for the working ontologist: modeling in RDF, RDFS and OWL, Amsterdam; Boston, Morgan Kaufmann Publishers/Elsevier.

\bibitem{WAL2008}Wal T.V.(2008), Understanding Folksonomy: Catalyzing Users to Enrich Information, O'Reilly Media, Inc.

\bibitem{thedata}The Dataverse Network, http://thedata.org

\bibitem{researchonresearch}Research on Research, DUKE University, http://www.researchonresearch.org

\bibitem{researchonresearch113}Research on Research, DUKE University, http://www.researchonresearch.org/?q=node/113

\bibitem{PIETROBON2004} Pietrobon R.,Guller U., Martins H., Menezes A.P., Higgins L.D. and Jacobs D.O.(2004), A suite of web applications to streamline the interdisciplinary collaboration in secondary data analyses. BMC Medical Research Methodology, 4, 29-29.

\bibitem{systemdynamics-biblio}System Dynamics Society,  http://www.systemdynamics.org/biblio/sdbibi.html

\bibitem{Yin2002}Yin R.K.(2002), Case Study Research: Design and Methods, Third Edition, Applied Social Research Methods Series, Vol 5, Sage Publications.

\bibitem{Marshall2006}Marshall C.; Rossman G. B.(2006), Designing qualitative research (4th ed.). Thousand Oaks, CA: Sage.

\bibitem{Silverman2004}Silverman P.D. (2004), Qualitative Research: Theory, Method and Practice, Sage Publications Ltd.

\bibitem{Laws2004}Laws K.; Mcleod R. (2004), Case Study and Grounded Theory : Sharing some Alternative Qualitative Research Methodologies with Systems Professionals. 22nd International Conference of the System Dynamics Society,Oxford, England, The System Dynamics Society.

\bibitem{Luna-Reyes2003}Luna-Reyes L.F.; Lines A.D. (2003), Collecting and analyzing qualitative data for system dynamics: methods and models. System Dynamics Review, 19, 271-296.

\bibitem{LINSTONEandTUROFF2002}Linstone H.A.;Turoff M. (2002), The Delphi Method Techniques and Applications, http://is.njit.edu/pubs/delphibook/delphibook.pdf

\bibitem{SOPDELPHI}SOP for video conference-based, real-time Delphi method, https://docs.google.com/document/d/1-jmu-IoF3mNHyY2mb2Qbq3P8xreF4kAWmas-NK5plcs/edit?hl=pt-BR

\bibitem{LEISCH2002}Leisch F. (2002),Sweave: Dynamic generation of statistical reports literate data analysis. Compstat 2002 - Proceedings in Computational Statistics, pages 575-580. Physica Verlag, Heidelberg. 

\bibitem{IMAI2006}Imai K.; King G. and Lau O. (2006), Zelig: Everyones Statistical Software, http://gking.harvard.edu/ zelig

\bibitem{r-project}The R Project for Statistical Computing, http://www.r-project.org

\bibitem{Moon2000}Moon J.; Sproull L. (2000), Essence of Distributed Work: The Case of the Linux Kernel. 

\bibitem{GitHub} GitHub, https://github.com

\bibitem{git}GitHub, https://git.wiki.kernel.org/index.php/GitSvnComparsion

\bibitem{subversion}Subversion, http://subversion.tigris.org

\bibitem{vensim}Ventana System Inc,  http://www.vensim.com

\bibitem{iseesystems}Stella/Ithink,  http://www.iseesystems.com

\bibitem{powersim} Powersim Software,  http://www.powersim.com

\bibitem{systems-thinking}Archetypes Interaction Structures of the Universe, http://www.systems-thinking.org/arch/arch.htm archdg 

\bibitem{MedModel}  Van den Belt M. and  Dietz T. (2004), Mediated Modeling A system dynamics approach to environmental consensus building. Island Press.

\end{thebibliography}
\end{document}